\documentclass{aa}

\usepackage{txfonts}
\usepackage{natbib}
\bibpunct{(}{)}{;}{a}{}{,} 

\usepackage{graphicx}
\graphicspath{{converted_graphics/}}

\def \s {3EG~J1835+5918}
\def \sagile {AGL~J1836+5926}
\def \aa {A$\&$A$\;$}
\def \agile {AGILE }

\def \degmark{^\circ}

\def \phcmsec{\hbox{photons cm$^{-2}$ s$^{-1}$}}

\def \arcmin {\hbox{$^\prime$}}
\def \arcsec {\hbox{$^{\prime\prime}$}}

\def \enmev {$E > 100 \: MeV$}

\begin{document}

\title{Long-term AGILE monitoring of the puzzling gamma-ray source 3EG J1835+5918 }

\date{}          
\authorrunning {A.Bulgarelli et al.}
\titlerunning { Long-term AGILE monitoring of the puzzling gamma-ray source 3EG J1835+5918 }

\author{A.~Bulgarelli\inst{1},
  M.~Tavani\inst{2,5}, P.~Caraveo\inst{4}, A.W.~Chen\inst{3,4},
  F.~Gianotti\inst{1}, M.~Trifoglio\inst{1}, M.~Marelli\inst{4},
  A.~Argan\inst{2}, G.~Barbiellini\inst{6}, F.~Boffelli \inst{9,15}, P.~W.~Cattaneo \inst{9},
  V.~Cocco\inst{2},
  E.~Costa\inst{1}, F.~D'Ammando\inst{2,5},
  E.~Del Monte\inst{2}, G.~De Paris\inst{2},
  G.~Di Cocco\inst{1}, I.~Donnarumma\inst{2},
  Y.~Evangelista\inst{2},
  M.~Feroci\inst{2}, M.~Fiorini\inst{4},
  T.~Froysland\inst{2,5},
  F.~Fuschino\inst{3}, M.~Galli\inst{7},
  A.~Giuliani\inst{4}, C.~Labanti\inst{1}, I.~Lapshov\inst{2}, F.~Lazzarotto\inst{2},
  P.~Lipari\inst{8}, F.~Longo\inst{6},
  M.~Marisaldi\inst{1}, S.~Mereghetti\inst{4},
  A.~Morselli\inst{10}, L.~Pacciani\inst{2},
  A.~Pellizzoni\inst{4}, F.~Perotti\inst{4},
  G.~Piano\inst{2,5},
  P.~Picozza\inst{5,10}, M.~Prest\inst{11},
  G.~Pucella\inst{2}, M.~Rapisarda\inst{12}, A.~Rappoldi \inst{9},
  P.~Soffitta\inst{2},  A.~Trois\inst{2},
  E.~Vallazza\inst{6}, S.~Vercellone\inst{4}, V.~Vittorini\inst{2,3},
  A.~Zambra\inst{4}, D.~Zanello\inst{8},
  P.~Giommi\inst{13},
  C.~Pittori\inst{13}, F.~Verrecchia\inst{13},
  P.~Santolamazza\inst{13},
  D.~Gasparrini\inst{13}, S.~Cutini\inst{13},
  S.~Colafrancesco\inst{13}, L.~Salotti\inst{11}
}

\institute{$^1$INAF/IASF--Bologna, Via Gobetti 101, I-40129 Bologna, Italy \\
$^2$INAF/IASF--Roma, Via del Fosso del Cavaliere 100,
  I-00133 Roma, Italy \\
$^3$CIFS--Torino, Viale Settimio Severo 3, I-10133, Torino, Italy \\
$^4$INAF/IASF--Milano, Via E.~Bassini 15, I-20133 Milano, Italy \\
$^5$Dip. di Fisica, Univ. ``Tor Vergata'', Via della Ricerca
  Scientifica 1, I-00133 Roma, Italy \\
$^6$Dip. di Fisica and INFN Trieste, Via Valerio 2, I-34127 Trieste, Italy\\
$^7$ENEA--Bologna, Via Biancafarina 2521, I-40059 Medicica (BO),
  Italy\\
$^8$INFN--Roma ``La Sapienza'', Piazzale A. Moro 2, I-00185 Roma,
  Italy\\
$^9$INFN--Pavia, Via Bassi 6, I-27100 Pavia, Italy\\
$^{10}$INFN--Roma ``Tor Vergata'', Via della Ricerca Scientifica 1,
  I-00133 Roma, Italy\\
$^{11}$Dip. di Fisica, Univ. dell'Insubria, Via Valleggio 11,
  I-22100 Como, Italy\\
$^{12}$ENEA--Roma, Via E. Fermi 45, I-00044 Frascati (Roma), Italy\\
$^{13}$ASI--ASDC, Via G. Galilei, I-00044 Frascati (Roma), Italy\\
$^{14}$ASI, Viale Liegi 26 , I-00198 Roma, Italy\\
$^{15}$Dipartimento di Fisica Nucleare e
Teorica, Universit\'a di Pavia, Via Bassi 6, Pavia, I-27100, Italy
}

\offprints{A. Bulgarelli, \email{bulgarelli@iasfbo.inaf.it} }
\date{received; accepted}

\abstract{ We present the AGILE gamma-ray observations of the
field containing the puzzling gamma-ray source 3EG J1835+5918.
This source is one of the most remarkable unidentified EGRET
sources. } { An unprecedentedly long AGILE monitoring of this source yields important
 information on the positional error box, flux
evolution, and spectrum. }{3EG J1835+5918 has been in the AGILE
field of view several times in 2007 and 2008 for a total observing
time of 138 days from 2007 Sept 04 to 2008 June 30 encompassing
several weeks of continuous coverage.}{With an exposure time 
approximately twice that of EGRET, AGILE confirms the existence
of a prominent gamma-ray source (\sagile) at a position consistent
with that of EGRET, although with a remarkably lower  average flux value for
photon energies greater than 100 MeV.
{A 5-day bin temporal analysis of the whole data set of
\sagile~ shows some evidence for variability of the gamma-ray flux. 
}
The source spectrum
between 100 MeV and 1 GeV can be fitted with a power law with
photon index in the range 1.6-1.7, fully consistent with the EGRET
value.} { The faint X-ray source RX J1836.2+5925 that has been
proposed as a possible counterpart of \s$\;$ is well within the AGILE
error box. Future continuous monitoring (both by AGILE and GLAST)
is needed to
confirm the gamma-ray flux variability and to
unveil the source origin, a
subject that is  currently  being pursued through a multiwavelength
search for counterparts. }

\keywords{ gamma-rays: observation;}


\maketitle

\section {Introduction}

The gamma-ray source 3EG J1835+5918 (also known as GEV
J1835+5921 and GRO J1837+59) is one of the most puzzling
high-energy sources in the sky, and it has been the subject of
considerable interest since its discovery by EGRET
\citep{Nolan_1994}. This source was catalogued among
the brightest high-latitude unidentified sources
with an average  gamma-ray flux above 100 MeV of $
\Phi_{EGRET}  = (60 \pm 4) \, 10^{-8} \,$ \phcmsec,
and  a hard gamma-ray spectrum of photon index
$\gamma = 1.69 \pm 0.07$ \citep{Hartman_1999}. Owing to the spectral hardness
(similar to that of the Vela pulsar), the
source flux near 1 GeV is expected to be about half of the Crab value.
The EGRET error box is
well-determined, resulting in an
error radius of 8\arcmin at the $95 \% $ confidence level.

The source \s$\;$ is located $25^{\circ}$ off the Galactic plane and
therefore is not significantly affected by the Galactic diffuse
gamma-ray emission. EGRET pointed at this high-galactic latitude
region several times \citep{Hartman_1999}. The resulting
position, gamma-ray lightcurve, variability, and spectrum have been
discussed in several papers: \citep{McLaughlin_1996, Nolan_1996, Reimer_2001}.
The analysis of the whole EGRET dataset produces a gamma-ray flux
database consistent with being constant \citep{Hartman_1999, Reimer_2001}.
However, it is important
to notice that a claim for variability of \s$\;$ was indicated by Nolan et al. (1996).

The relatively strong gamma-ray flux, the hard spectrum, and the
lack of an obvious blazar counterpart in the EGRET error box of \s~
spurred considerable interest from several observing groups.
After extensive X-rays, radio, and optical coverage of the EGRET
error box  \citep{Mirabal_2000, Reimer_2001, Totani_2002, Halpern_2002} 
the search for possible
counterparts singled out RX~J1836.2+5925, a relatively faint soft
X-ray source with no radio or optical emission: a set of
characteristics similar to those of middle-aged radio-quiet
neutron stars \citep{Caraveo_1996}. Thus, on the basis of the
gamma-ray properties of \s, coupled with those of RX~J1836.2+5925,
the sources was named \textit{the next Geminga} by \citep{Halpern_2002}
owing to its Geminga-like phenomenology \citep{Bignami_1996}. A
search for X-ray pulsation, with period values ranging from 1 ms
to 10 s, yielded a 35\% upper limit to the source-pulsed fraction
\citep{Halpern_2007}. In view of the much lower pulsed fractions
measured for all known middle-aged neutron stars, such a result
did not weaken this tentative identification.

Taking advantage of the AGILE unprecedented exposure gathered during the
science verification phase and early pointings of the  Cycle-1
program,  in this Letter we are now able to assess the source
characteristics anew. Section 2 describes the AGILE-GRID observations
and main results, while their implications are discussed in
Section 3.

\section{The AGILE GRID observations of \s}

The AGILE istrument \citep{Tavani_2008} is composed of three
detectors: a Tungsten-Silicon Tracker (ST) \citep{Barbiellini_2002, Prest_2003}, 
with a large field of view ($\sim 2.5 $~sr),
optimal time resolution and  angular resolution, and good sensitivity;
a Silicon-based X-ray detector, Super-AGILE (SA) \citep{Feroci_2007} for  imaging
in the energy range 18~keV - 60~keV and a CsI(Tl) Mini-Calorimeter
(MCAL) \citep{Labanti_2006} that detects gamma-rays or particle
energy depositions between 300~keV and 100~MeV. The 
ST and MCAL form the AGILE  Gamma-Ray Imaging Detector (GRID)
for observations in the gamma-ray energy range 30~MeV - 50~GeV. The instrument is
completed by an anti-coincidence (AC) system
\citep{Perotti_2006}, made with plastic scintillator layers, for
the rejection of charged particles,  and by an efficient trigger logic for
gamma-ray and X-ray data acquisition \citep{Tavani_2008}.

\begin {table*}[!htb]
\caption {\em{AGILE observations of \s.}}
\label{table_1}
\renewcommand{\arraystretch}{1.2} 
\begin{tabular}{@{}llllllllcclll}
\hline
\textbf {Viewing period} & \textbf{ l } & \textbf{ b } & \textbf{ Start (UTC) } &
\textbf{ End (UTC) } &  \textbf { Off-axis angle } & $F_\gamma$ & $\Delta F_{\gamma}$ &
Counts & $\sqrt(TS)$ \\
\hline
OB2300  & 134.882 & 11.822  & 2007-09-04 12:00 & 2007-09-12 12:00 & 50 & $<$ 40 & & $<$ 13   & 1.8\\
OB4610 + OB4630  & 143.365 & 41.588  & 2007-10-24 08:00 & 2007-11-01 12:00 & 50  & 81 & 18  & 37 $\pm$ 10 & 4.4 \\
OB4800  & 69.594 & 4.623 & 2007-11-02 12:00  & 2007-12-01 12:00 & 27  & 40 &  6 & 99  $\pm$ 16 & 7.7  \\
OB4900  & 88.815 & 9.928 & 2007-12-01 12:00 & 2007-12-05 09:00 & 15  & 39 & 14 & 18  $\pm$ 5 & 3.8  \\
OB4910 + OB4920  & 85.119 & -9.416 & 2007-12-05 09:00 & 2007-12-16 12:00 & 38  & 48 & 12 & 44  $\pm$ 10 & 5.2  \\
OB5210  & 77.309 & 40.628 & 2008-02-09 09:00 & 2008-02-12 12:00 & 20 & 60 & 19 & 21  $\pm$ 7 & 4.2   \\
OB5600  & 53.039 & 6.474 & 2008-04-10 12:00  & 2008-04-30 12:00 & 40 & 40 & 10 & 57  $\pm$ 13 & 5.0  \\
OB5700  & 104.852  & 35.439 & 2008-04-30 12:00  & 2008-05-10 12:00 & 30  & 49 & 11 & 49  $\pm$ 10 & 6.2  \\
OB5800  & 74.05  & 0.273  & 2008-05-10 12:00  & 2008-06-09 12:00  & 30  & 50  &  9  & 129  $\pm$ 18  & 9.6  \\
OB5820  & 93.6  & -1.16  &  2008-06-15 12:00 & 2008-06-30 12:00 & 30  & 25 &  8 & 37  $\pm$ 11 & 4.2 \\
\hline
\end{tabular}
\end{table*}

 Data analysis was  performed with the BUILD 15 of the AGILE
Standard Pipeline, publicly available at the ASI Data Center web
site (http://agile.asdc.asi.it/). The data were produced
starting from the Level-1 data. The events collected during the
passage in the South-Atlantic Anomaly and the Earth albedo
background were consistently rejected. The GRID event
direction were reconstructed by a Kalman filter technique. To
reduce the particle background contamination, we selected 
with the GRID filter $\rm FT3ab\_2$  only events flagged as
confirmed $\gamma$-ray events (\textit{G} class events, corresponding to
a sensitive area of $\sim 300 \: cm^2$   at 100 MeV).
All the fluxes reported in this Letter have been consistently
checked and scaled with their corresponding errors 
using the ratio $R = \frac{\Phi(Vela)_{EGRET}}{\Phi(Vela)_{AGILE}}$
computed for similar off-axis angles, with $\Phi(Vela)$ the
gamma-ray flux above 100 MeV reported by EGRET and measured by
AGILE.
Scaling our fluxes measured for \s$\;$ to the Vela PSR
measurements by AGILE and EGRET also takes systematic
effects into account (estimated to be of the order of the 10\% of the fluxes
reported here). 

All AGILE observations of the field containing the \s$\;$ are listed
in Table \ref{table_1}. The columns report the viewing periods name
(as reported in the ASI Data Center web site), the galactic coordinates of the centre of
the field of view, the start and end of the observation in UTC, the average source
off-axis angle, the
{\it period-averaged} flux $F_{\gamma}$ (\enmev) in \phcmsec (if $\sqrt{TS}<3$ a $2-\sigma$ upper limit
is reported), the $1-\sigma$ statistical  and systematic uncertainty of the flux,
the number of counts, and in the last column the statistical significance of the likelihood ratio test.

AGILE counts, exposure, and galactic background maps were generated
with a bin size of $0.25^{\circ} \times 0.25^{\circ}$ for \enmev~
to compute the source flux and its evolution, while maps with a bin
size of $0.1^{\circ} \times 0.1^{\circ}$ were used to optimize
the source position using a likelihood analysis method
\citep{Mattox_1996a}. Our analysis was performed over a region of $10\degmark$
radius.

\subsection{The gamma-ray source positioning}

\begin{figure}[!htb]
\centering
\includegraphics[width=9 cm]{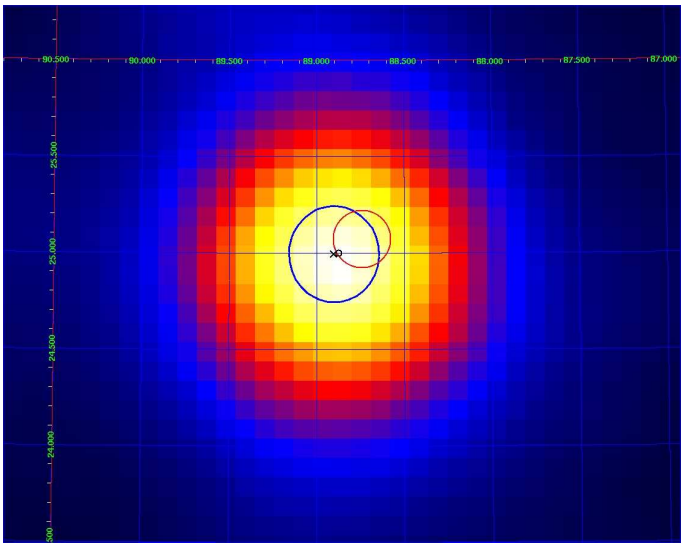}
\centering \caption { {\itshape Gaussian-smoothed counts map with
squared scale of \sagile$\;$ with photons of \enmev. The red curve is
the EGRET 95$\%$ error ellipse, the blue
curve is the AGILE 95$\%$ statistical plus systematic error ellipse,
the cross is the AGILE 95$\%$ error ellipse barycentre, and the
black circle is the position of the X-ray source
 RX J1836.2+5925. } } \label{fig_position}
\end{figure}

To determine the most likely location of the gamma-ray
source, we used only the best viewing periods, i.e. those in
which the source was within $50^{\circ}$ from the instrument
pointing direction. For \enmev, the best position computed by the
likelihood analysis
 is $l=88.90^{\circ} \; b=24.99^{\circ}$, corresponding to
 $\alpha=18\degmark 36\arcmin 20\arcsec.11$, $\delta=59\degmark 26\arcmin 42\arcsec.3$.
We name this source \sagile. Figure
\ref{fig_position} shows a Gaussian-smoothed counts map of the
source with our 95$\%$ statistical and systematic error ellipse  with
semiaxis $a=0.265^{\circ}$ and $b=0.200^{\circ}$. To conservatively
take both  systematic and statistical effects into account,
we obtain a 95$\%$ confidence level radius
by linearly adding a systematic error of $0.1^{\circ}$ to the 95$\%$ statistical error ellipse.

The AGILE 95$\%$ error ellipse is  consistent with
the error box of \s$\;$ \citep{Mattox_2001}, and the distance between AGILE and EGRET centroids is $\sim
10.2$ \arcmin.
The distance between the position of the X-ray source RX J1836.2+5925 and the
AGILE 95$\%$ maximum likelihood contour level barycentre is $\sim
1.4$ \arcmin.
The source detection significance, as derived from a maximum
likelihood analysis, is 16.7-$\sigma$.
A total of 499 photons with $E > 100 \: $~MeV have been collected.
The source flux, averaged over the whole AGILE dataset,
is $\Phi_{AGILE} = (38.7 \pm 3.0) \times 10^{-8}$ \phcmsec $\:$,
 which turns out to be approximately  60$\%$ less than 
the value reported in the 3rd EGRET Catalogue.

\subsection{The gamma-ray spectrum}

Our standard GRID spectral analysis was performed
using 3 energy bins, namely 100-200 MeV, 200-400 MeV, 400-1000
MeV.
Photons below 100 MeV  and  above 1 GeV have not been used
in the analysis reported here.
 The analysis was obtained
by summing the observation periods OB5600, OB5700, and OB5800 (see
Table \ref{table_1}).  A power-law model fit to the data
yields a  photon index $\gamma = 1.65 \pm 0.22$.

\subsection{The gamma-ray lightcurve and variability analysis}

The search for the variability of the gamma-ray emission of \s$\;$
produced somewhat inconclusive results. Using the first 
EGRET viewing periods, McLaughlin et al. (1996) and
Nolan et al. (1996) classified  \s$\;$ as a possibly variable source.
On the other hand,  using the whole EGRET data set Reimer et al. (2001)
and Nolan et al. (2003) 
did not confirm the previous hint of variability. 
(We note that none of the previous EGRET variability studies used
the same 5-day intervals adopted in this Letter.)

\begin{figure*}[!ht]
\centering
\includegraphics [width=19 cm] {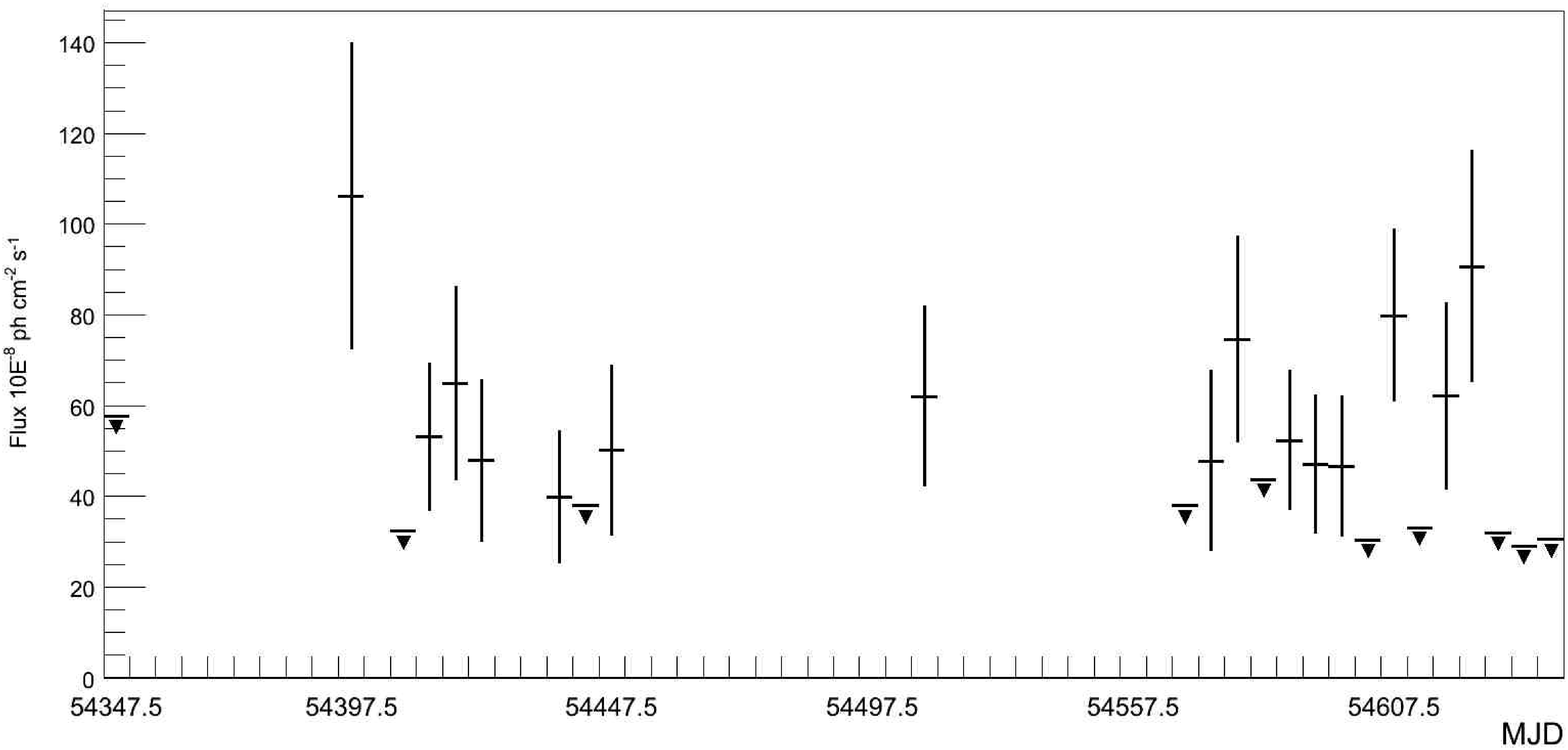}
\centering \caption { {\itshape AGILE gamma-ray light-curve of \sagile~
for photons above 100 $MeV$ with a  5-day temporal bin.
}
} \label{fig_LC_J1835_D5}
\end{figure*}

To study the source temporal behaviour,
each AGILE viewing period was first analysed to determine the parameters
of the Galactic diffuse radiation model and of the isotropic diffuse
intensity level.
The  source flux density was then
estimated independently for each 5-day temporal bin  with the
gamma-ray model parameters fixed at the values obtained in the first step.
Figure \ref{fig_LC_J1835_D5} shows the  \sagile~ gamma-ray flux lightcurve
during the period September 2007 to June 2008 for photons with \enmev.

We fit  the AGILE GRID fluxes to a constant value (the weighted
mean of the average flux values), and  computed the $V$ variability
coefficient for the AGILE observations \citep{McLaughlin_1996}. The
2-$\sigma$ upper limits (UL) were properly treated, assigning a
value and a $\sigma$ equal to $UL/2$ \citep{Torres_2001a}. A value
of $V > 1$
is an indication of variability within the observing period.
The AGILE observation block durations of Table \ref{table_1}
span a non-homogeneous range of timescales from 1 week to more
than a month. To study the flux variability of \sagile~ on
a timescale of a few days, we chose a temporal bin size of 5 days
as a good compromise between statistics and variability
characteristics.
 Using a 5-day bin size, we obtain $\chi^{2} = 38.34$ for
24 degrees of freedom and a value of $V = 1.495$. We can exclude
the flux being constant at the $96.80\%$ level.  For a temporal
bin of 4 days we obtain $V\simeq1$. For  temporal bins of 3
days or shorter, there are too many upper limits during the whole
exposure time. For temporal bins of more than 5 days, we obtain $V<1$, 
as expected for  averaging a signal potentially variable on
a timescale of a few days. 

We note that, during the long
uninterrupted coverage obtained in Nov-Dec 2007 and in April-June
2008, the gamma-ray source  \sagile$\;$ was not detectable  (failing to
reach the 3-$\sigma$ threshold) for about half of the observing
time. We also determined hard X-ray upper limits for \sagile$\;$ in the energy range
18-60 keV from an analysis of Super-AGILE data.
A 3-sigma upper limit of 10 mCrab was obtained for the OB5700.

\subsection{EGRET vs AGILE}

After one year of
operations, AGILE has doubled
the EGRET exposure time of \s, confirming the source position
and spectral shape and  showing evidence of variability.
A comparison of the AGILE and EGRET observations  of  3EG J1835+5918
is reported in Table \ref{table_2}.
With a few exceptions, the EGRET pointings were generally one-week
long \citep{Reimer_2001}, and they were not optimally suited to
tracing an emission variable on week to month long timescales.
On the other hand, \sagile$\;$ has been continuously in the
AGILE field of view for more than two months in April-June 2008, in addition
to several weeks scattered between September 2007 and December
2007. It is too early to say whether the apparent variations are erratic or 
follow some kind of cyclic pattern. Longer, continuous
exposures (both with AGILE and GLAST) are needed to clarify this
crucial point.

\begin {table}[!htb]
\caption {\em{Comparison of total exposure and number of photons gathered by EGRET and AGILE from 3EG J1835+5918 for $E > 100 \: MeV$.}} \label{table_2}
\renewcommand{\arraystretch}{1.2} 
\begin{tabular}{@{}lllllllllllll}
\hline
\textbf {} & \textbf{ $cm^2 \: s \: sr \diagup 0.5\degmark$ bin }  & \textbf{ Photons $E > 100 \: MeV$ } \\
\hline
EGRET  & 52900  & 452 \\
AGILE  & 118325 & 499 \\
\hline
\end{tabular}
\end{table}

\section{Discussion and conclusions}

The AGILE observations and monitoring of \sagile$\;$ adds relevant
information about this puzzling source. Postponing an account of our
multifrequency observations of the region to a forthcoming paper, we briefly outline
here a few important points regarding the search for a counterpart.
We confirm a point already noticed by other authors, i.e., the absence of
a blazar or any other relatively bright
radio  sources  in the field containing the AGILE (and EGRET) error box.
 The analysis
of  Mattox et al. (2001) shows a 100 mJy source (B1834+5904)
positioned at the 99.5$\%$ probability contour ($12'.7$ from the
EGRET centroid position and $21'.8$ from the AGILE centroid position). By using
the NED, SIMBAD, and  the  Massaro et al. (2007) catalogues, we confirm the
absence of a radio-loud blazar  in the revised AGILE error box.

Outside this error box, we
notice the existence of  RGB J1841+591, a BL Lac type object
positioned at $43'.5$ from  the AGILE centroid position.
Even though occasional contributions from this object cannot
be excluded in our analysis of \agile, we emphasise that the AGILE
integrated flux positioning at the 95\% contour level is clearly
not consistent with any substantial contribution from
RGB~J1841+591. 

We also notice that the faint X-ray source RX J1836.2+5925
that was proposed as a possible counterpart is well within
the AGILE error box of \sagile.
A hint of variability of this source was noticed when comparing two HRI ROSAT
observations taken almost three years apart \citep{Mirabal_2001}, prompting
  a claim for the
discovery of a new class of compact gamma-ray sources \citep{Mirabal_2000}.
However, subsequent
Chandra X-ray observations showed a practically constant X-ray flux
of RX J1836.2+5925, and weakened the variability claim
in favour of a scenario encompassing a constant Geminga-like
source.

We note here that our group observed 
the region containing RX J1836.2+5925 with the SWIFT XRT and XMM-Newton
during the period May-June 2008.
Analysis of these data and a full discussion of the \sagile$\;$ counterpart problem will
be presented elsewhere.

Future continuous monitoring (both by AGILE and GLAST) is needed to
 confirm the gamma-ray flux variability and to unveil the source origin.


\end{document}